\documentclass[aps,showpacs,nofootinbib,superscriptaddress]{revtex4}
\usepackage{epsf}
\usepackage{graphicx}
\usepackage{amsmath}
\begin{document}

\title{Theoretical uncertainties on quasielastic charged--current
  neutrino--nucleus cross sections} \author{M.~Valverde}

\author{J.~E.~Amaro} \author{J.~Nieves}
\affiliation{Departamento de F\'\i sica At\'omica, Molecular y Nuclear
\\ Universidad de Granada, E-18071 Granada, Spain}

\begin{abstract}
We estimate the theoretical uncertainties of the model developed in
Phys. Rev. {\bf C70} 055503 for inclusive quasielastic
charged--current neutrino--nucleus reactions at intermediate
energies. Besides we quantify the deviations of the predictions of
this many body framework from those obtained within a simple Fermi gas
model.  A special attention has been paid to the ratio
$\sigma(\mu)/\sigma(e)$ of interest for experiments on atmospheric
neutrinos. We show that uncertainties affecting this ratio are likely
smaller than 5\%
\end{abstract}

\maketitle

\section{Introduction}

Recently an interest in neutrino scattering off nuclei has raised
because of its implications in the experiments on neutrino
oscillations based on large Cerenkov detectors.  Atmospheric neutrino
oscillations are dominantly due to $\nu_\mu \to \nu_\tau$ flavor
mixing, which has been already confirmed by the Super-Kamiokande
experiments~\cite{fukuda}.  Future experiments aim at observing
sub-leading effects on top of the dominant one~\cite{procs}, which are
due to atmospheric $\nu_e$ flavor oscillations. Observability of these
effects depends strongly on the assumed experimental and theoretical
systematic uncertainties. The most simple model regarding these
effects supposes that these oscillations are controlled by neutrino
masses (through their mass squared differences $\Delta m^2$) and by a
$3\times 3$ unitary mixing matrix similar to the CKM matrix in the
quark sector. Thus we are provided, besides the mass squared
differences, with 3 (mixing angles) + 1 (CP-violating phase)
parameters not fully determined yet.  To reduce the error size on
these parameters requires an improvement in already existing neutrino
detectors. These experiments use the nuclei in water as target for the
incoming neutrinos, so a nuclear reaction model is needed for the data
analyses.  One of the main sources of systematical error in these
experiments is precisely the cross sections for quasielastic (QE)
neutrino scattering \cite{kajita}.  An accurate theoretical framework
for the neutrino--nucleus dynamics, and a reliable estimate of the
uncertainties affecting its predictions will definitely help to have a
close control over the systematics affecting the oscillation
parameters.

There is a rich recent literature on the Charged Current (CC)
neutrino-nucleus reactions in the QE region at intermediate
energies~\cite{HT00}--\cite{Amaro:2004bs}. However,  a systematic
analysis of the theoretical uncertainties affecting the predictions of
these works is often missing. We will use here the Many Body
Framework (MBF) developed in Ref.~\cite{Nie04}\footnote{This model has
been recently extended to the Neutral Current (NC) sector and to the
study of nucleon knock-out reactions induced by
neutrinos~\cite{Nie06}.}. Starting from a Local Fermi Gas (LFG)
picture of the nucleus, which automatically accounts for Pauli
blocking, several nuclear effects are taken into account in that
scheme: {\it i)} a correct energy balance, using the experimental
$Q-$values, is enforced, {\it ii)} Coulomb distortion of the charged
leptons is implemented by using the so called ``modified effective
momentum approximation'', {\it iii)} medium polarization (RPA),
including $\Delta-$hole degrees of freedom and explicit pion and rho
exchanges in the vector--isovector channel of the effective
nucleon--nucleon force, and Short Range Correlation (SRC) effects are
computed, and finally {\it iv)} the nucleon propagators are dressed in
the nuclear medium, which amounts to work with a LFG of interacting
nucleons and it also accounts for reaction mechanisms where the gauge
boson, $W^+$ or $W^-$, is absorbed by two nucleons.  This model is a
natural extension of previous studies on electron~\cite{GNO97},
photon~\cite{CO92} and pion~\cite{OTW82,pion} dynamics in nuclei. Even
though the scarce existing CC data involve very low nuclear excitation
energies, for which specific details of the nuclear structure might
play an important role, the model of Ref.~\cite{Nie04} provides one of
the best existing combined descriptions of the inclusive muon capture
in $^{12}$C and of the $^{12}$C $(\nu_\mu,\mu^-)X$
and $^{12}$C $(\nu_e,e^-)X$ reactions near threshold.  Inclusive muon
capture from other nuclei is also successfully described by the
model. Above 80 or 100 MeV of energy transferred to the
nucleus, this MBF leads also to excellent results for the
$(e,e^\prime)$ inclusive reaction in nuclei, not only in the QE
peak, but also in the $\Delta$
and the dip (located between the QE and the $\Delta$
peaks) regions~\cite{GNO97}\footnote{Data in $^{12}$C, $^{40}$Ca and
$^{208}$Pb of differential cross sections for different electron
kinematics and split into longitudinal and transverse response
functions are successfully described. }. It also successfully describes  the
absorption of real photons by nuclei in this energy regime~\cite{CO92}.

In this work we pay special attention to the source and size of the
theoretical uncertainties affecting the predictions of
Ref.~\cite{Nie04}. Firstly  we assign to each of the main inputs of the
model a reliable uncertainty. For the
experimentally determined parameters this obviously has to be
determined by the experimental error, while for the model dependent
parameters, we will assume theoretically founded sizes for their
errors.  Then we propagate the errors by means of a numerical
simulation, that is: we consider the input parameters to be
represented by uncorrelated Gaussian distributions, and by means of a
Monte Carlo (MC) simulation, we find for any observable predicted by
the model its derived probability distribution, which specific
features will determine its associated theoretical uncertainty.

Besides, there exist some systematic errors associated to the
validity of  the hypothesis in which the scheme of Ref.~\cite{Nie04}
is based. Those are harder to estimate and will be discussed   at
the end of this paper.

\section{Sources of theoretical  errors}
 
The main inputs of  the model of Ref.~\cite{Nie04} are:
\begin{itemize}

\item {\it Lepton and hadron masses, electro weak coupling constants},
which we will assume to be errorless.

\item {\it Neutrino-nucleon form factors.}

The neutrino-nucleon interaction is assumed to be of the $V-A$ type.
The vector form factors are related to the electromagnetic ones by
means of the isospin symmetry, which we will consider exact. For the
electromagnetic form factors we use the Galster
parameterization~\cite{Galster},
\begin{eqnarray}
F_1^N &=& \frac{G_E^N+\tau G_M^N}{1+\tau},  \\
 \mu_N F_2^N &=& \frac{G_M^N- G_E^N}{1+\tau},  \\
G_E^p &=& \frac{G_M^p}{\mu_p}=
 \frac{G_M^n}{\mu_n}=-(1+\lambda_n\tau)\frac{G_E^n}{\mu_n\tau} \\\nonumber
&=&\left(\frac{1}{1-q^2/M^2_D}\right)^2
\end{eqnarray}
with $\tau=-q^2/4M^2$, $M$ the nucleon mass, $N$ standing for $n$
(neutron) or $p$ (proton) and $\mu_N$, the corresponding nucleon
magnetic moment.  The axial part is ruled by the axial form factor
\begin{equation} 
G_A(q^2) = \frac{g_A}{(1-q^2/M_A^2)^2}
\end{equation}
The pseudoscalar $G_P$ form factor is related to the axial one by
means of the partially conserved axial current hypothesis
(PCAC).  The non--pole contribution to $G_P$, at $q^2=0$, is around 200 times
smaller~\cite{ADW} than the pole contribution taken into account by means of
PCAC. We will not explicitly consider  this source of systematics,
though we will assume an uncertainty for $g_A$ larger than that
generally used in the literature.

Thus, from these form factors we are left with four input
parameters: $M_D$, $\lambda_n$, $g_A$ and $M_A$, since we neglect any
type of uncertainty in the nucleon magnetic moments.  The Particle
Data Group (PDG) compiles several determinations of
$g_A(0)/g_V(0)$\footnote{We assume $g_V(0)=1$ according to the
conserved vector current hypothesis.} from neutron beta decay studies,
ranging most of them in the interval 1.25--1.27~\cite{pdg}, we will
adopt a conservative point of view and we will take here 1.26$\pm
0.01$, though the average error quoted by the PDG is more than three
times smaller. For the axial cutoff mass, we will assume 1.05 $\pm$
0.14 from the analysis of the $\nu d \to \mu^- pp$ reaction carried
out in Refs.~\cite{BNL1,BNL2}. Finally we will take a 5\% and 10\%
error on $M_D$ and $\lambda_n$, respectively. Parameters are compiled
in Table~\ref{tab:par}.

\item {\it The nuclear medium effective baryon--baryon interaction used in the
computation of the RPA effects.}

The effective nuclear interaction is included in the model starting
from an interaction between particle-hole (ph) pairs of the
Landau-Migdal type
\begin{equation}
V =  c_{0}\left\{f_{0}^{\prime}(\rho)\vec{\tau}_{1}\vec{\tau}_{2}
+g_{0}^{\prime}(\rho)
\vec{\sigma}_{1}\vec{\sigma}_{2}\vec{\tau}_{1}\vec{\tau}_{2}
\right\},
\end{equation}
for the isovector channels\footnote{The isoscalar channels do not
  contribute to CC induced reactions.}. The $\rho(r)$ dependence of $f_0^\prime$ is linearly parameterized
as~\cite{speth}
\begin{equation}
f_{0}^\prime(\rho (r))=\frac{\rho (r)}{\rho (0)} f_{0}^{\prime (in)}+
\left[ 1-\frac{\rho (r)}{\rho (0)}\right] f_{0}^{\prime (ex)}
\end{equation}
Thus, we have from the $\vec{\tau}\vec{\tau}$ channel
three parameters: $c_0$, $f_{0}^{\prime (in)}$ and $f_{0}^{\prime (ex)}$. 
In the  channel $\vec{\sigma}\vec{\sigma}\vec{\tau}\vec{\tau}$ we use an
interaction with explicit $\pi$ (longitudinal) and $\rho$ (transverse)
exchanges and then we  replace~\cite{OTW82}
\begin{equation}
c_0g_{0}^{\prime}(\rho) \vec{\sigma}_{1}\vec{\sigma}_{2}
\vec{\tau}_{1}\vec{\tau}_{2} \to \vec{\tau}_{1}\vec{\tau}_{2}
\sum_{i,j=1}^3\sigma_{1}^i \sigma_{2}^j V_{ij}^{\sigma\tau} \label{eq:lm}
\end{equation}
where
\begin{equation}
V_{ij}^{\sigma\tau} = \left (\hat{q}_{i}\hat{q}_{j}V_{l}(q)
+({\delta}_{ij}- \hat{q}_{i}\hat{q}_{j})V_{t}(q)\right) 
\end{equation}
$\hat{q} = \vec{q}/|\vec{q}\,|$ is an unitary vector parallel to the
transfered momentum and the strengths of the ph-ph
interaction in the longitudinal and transverse channel are given by
\begin{align}
V_t(q^0,\vec{q}) &= \frac{f^2}{m^2_\pi}\left\{
   C_\rho \left(\frac{\Lambda_\rho^2-m_\rho^2}{\Lambda_\rho^2-q^2 }\right)^2
\frac{\vec{q}{\,^2}}{q^2-m_\rho^2} + g^\prime_t(q)\right \}, \nonumber \\
V_l(q^0,\vec{q}) &= \frac{f^2}{m^2_\pi}\left\{
\left(\frac{\Lambda_\pi^2-m_\pi^2}{\Lambda_\pi^2-q^2 }\right)^2
\frac{\vec{q}{\,^2}}{q^2-m_\pi^2} + g^\prime_l(q)\right \}
\end{align}
Besides, $\Delta(1232)$ degrees of freedom are also taken into account
in this channel.  The ph--$\Delta$h and $\Delta$h--$\Delta$h effective
interactions are obtained from the interaction of Eq.~(\ref{eq:lm}) by
replacing $\vec{\sigma} \to \vec{S} $, $\vec{\tau} \to \vec{T} $,
where $\vec{S},\vec{T} $ are the spin, isospin $N\Delta$ transition
operators~\cite{OTW82} and $f\to f^*=2.13~f$. The SRC functions
$g^\prime_l$ and $g^\prime_t$ have a smooth
$q-$dependence~\cite{OTW82}, which is not considered here, since we
will explore only low and intermediate energies and momenta, and thus
we  take
$g^\prime_l(q)=g^\prime_t(q)=g^\prime$~\cite{Nie04,GNO97,pion}. Hence,
we end up with six additional parameters: $f$, $f^*$ $\Lambda_\pi$,
$C_\rho$, $\Lambda_\rho$ and $g^\prime$. We would like to point out
that all these nine parameters, which are used to parameterize the
medium effective baryon--baryon interaction, were adjusted long time
ago~\cite{OTW82,speth}, and since then have been successfully used in
several nuclear calculations at intermediate energies~\cite{GNO97}--
\cite{pion}. Here we have assumed uncorrelated Gaussian distributions
with relative errors of 10\%, for all these parameters (see
Table~\ref{tab:par}), except for the constant $c_0$, for which we have
not considered any type of uncertainty, because it always appears
multiplying the parameters $f_{0}^{\prime (in)}$ and $f_{0}^{\prime
(ex)}$ that have already 10\% error within our analysis.

\item {\it The nucleon self--energy ($\Sigma$).} 

The nucleon self--energy is used to dress the nucleon propagators in
the medium. Its real part modifies the free nucleon dispersion
relation, while the imaginary part takes into account two nucleon
absorption reaction channels. The result of it is a quenching of the
QE peak with respect to the simple ph excitation calculation and a
spreading of the strength, or widening of the peak. The integrated
strength over energies is not much affected though. Most of the effect
comes from the consideration of the real part of the nucleon
selfenergy, as it was first pointed out in
Ref.~\cite{Nie06}\footnote{See for instance solid (red) and dotted
(magenta) lines in Fig.4 of that reference.}.  The model of
Ref.~\cite{Nie04} uses the results of a semiphenomenological approach
developed by Fern\'andez de C\'ordoba and Oset in \cite{FO92}. The
nucleon selfenergy, $\Sigma (p^0 , \vec{p}\,; \rho)$, calculated in
this latter reference depends explicitly on the nucleon energy and
momentum~\cite{FO92} and leads to nucleon spectral functions in good
agreement with accurate microscopic approaches like the ones of
Refs.~\cite{Fa83}--\cite{Ra89}. Because in great part the model of
Ref.~\cite{FO92} is not entirely microscopical, it is hard to identify
its parameters. Here we have assumed a 10\% relative
error (Gaussian) for the nucleon selfenergy in the
medium. Ten percent error is a reasonable choice, since it safely
covers the existing differences  between the results of
Ref.~\cite{FO92} and the microscopic approaches of
Refs.~\cite{Fa83}--\cite{Ra89}. As discussed above, the change in the
nucleon dispersion relation is more important than the inclusion of
the small nucleon width in the medium, related to the quasielastic
channels, which will account for $W^\pm$ absorption by two
nucleons. Neglecting completely the imaginary part of the nucleon
selfenergy leads to a considerable reduction in computation time and given the
quality of this approximation, it has been used in all  MC
simulations performed to estimate the theoretical uncertainties of the
results of Ref.~\cite{Nie04}.

\item {\it Proton and Neutron matter densities.}

 Charge densities are taken
from \cite{ExpDens} and proton densities are deduced from them. On the
other hand, neutron densities are taken approximately
equal (but normalized to the number of neutrons) to the proton ones,
though small changes are considered, inspired by Hartree-Fock
calculations with the density-matrix expansion~\cite{Ne75} and
corroborated by pionic atom data~\cite{GNO92}. In this work we will
present results for oxygen, carbon and argon, for the first two we use
a modified harmonic oscillator distribution MHO (parameters can be
found in Table~I of Ref.~\cite{Nie04}), while for argon we use a
two-parameter Fermi distribution (parameters can be found in Table~I
of Ref.~\cite{Nie06}). We take a 5\% relative error (Gaussian uncorrelated)
for all parameters, which is about one order of magnitude larger than
the quoted errors for charge density parameters, but that  safely
covers uncertainties related to the neutron distributions and to the
procedure of taking out the the finite size of the proton and  
neutron particles.

\end{itemize}

\begin{figure}
\begin{center}
\makebox[0pt]{\includegraphics[height=20cm]{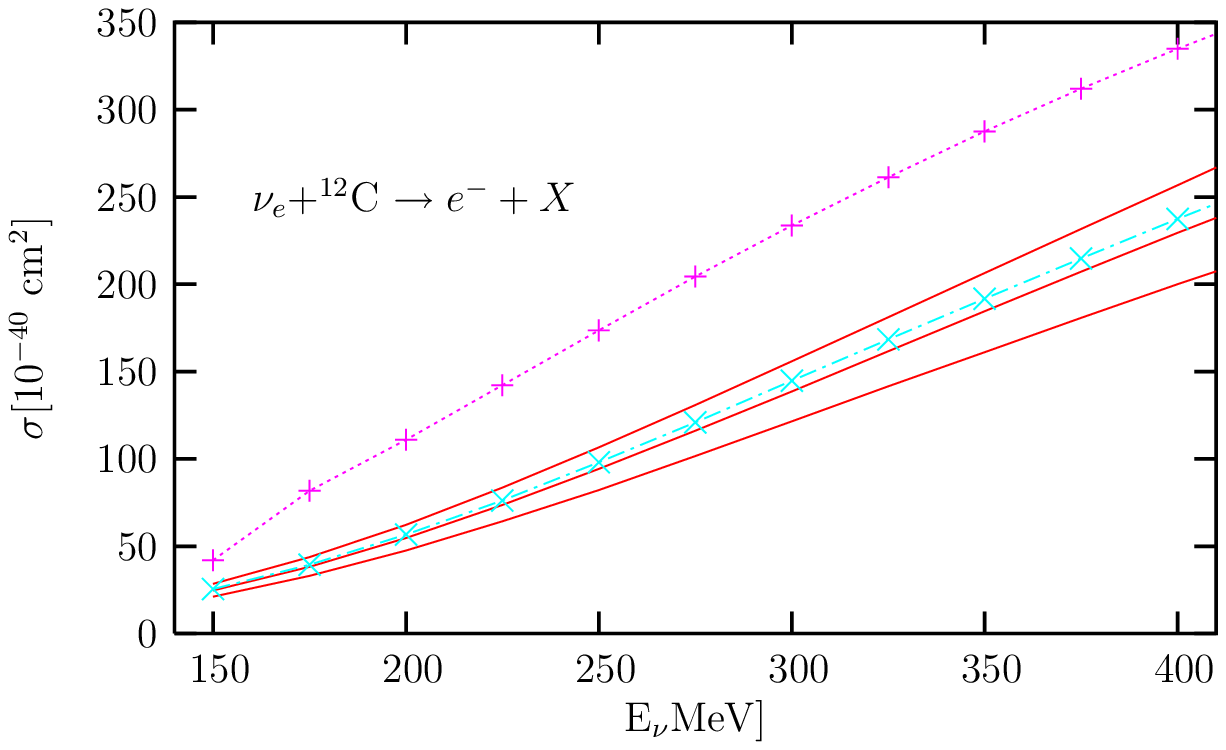}\hspace{-5cm}\includegraphics[height=20cm]{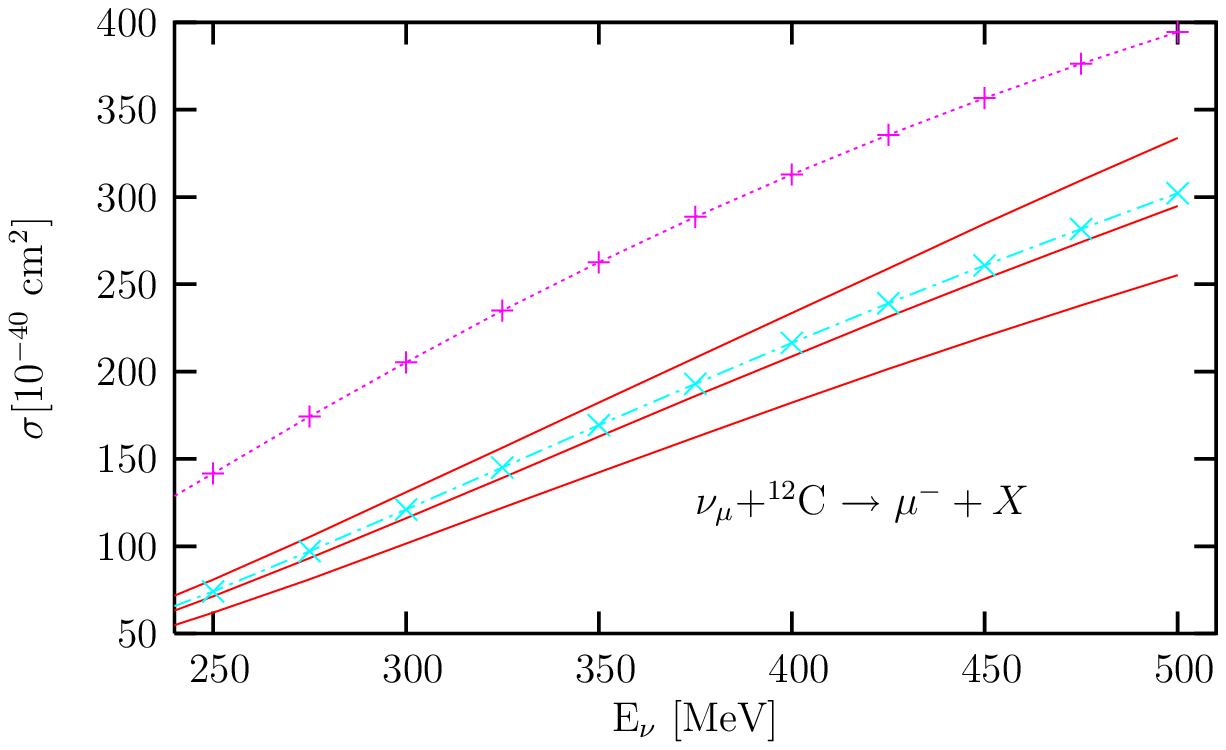}}\\[-14cm]
\makebox[0pt]{\includegraphics[height=20cm]{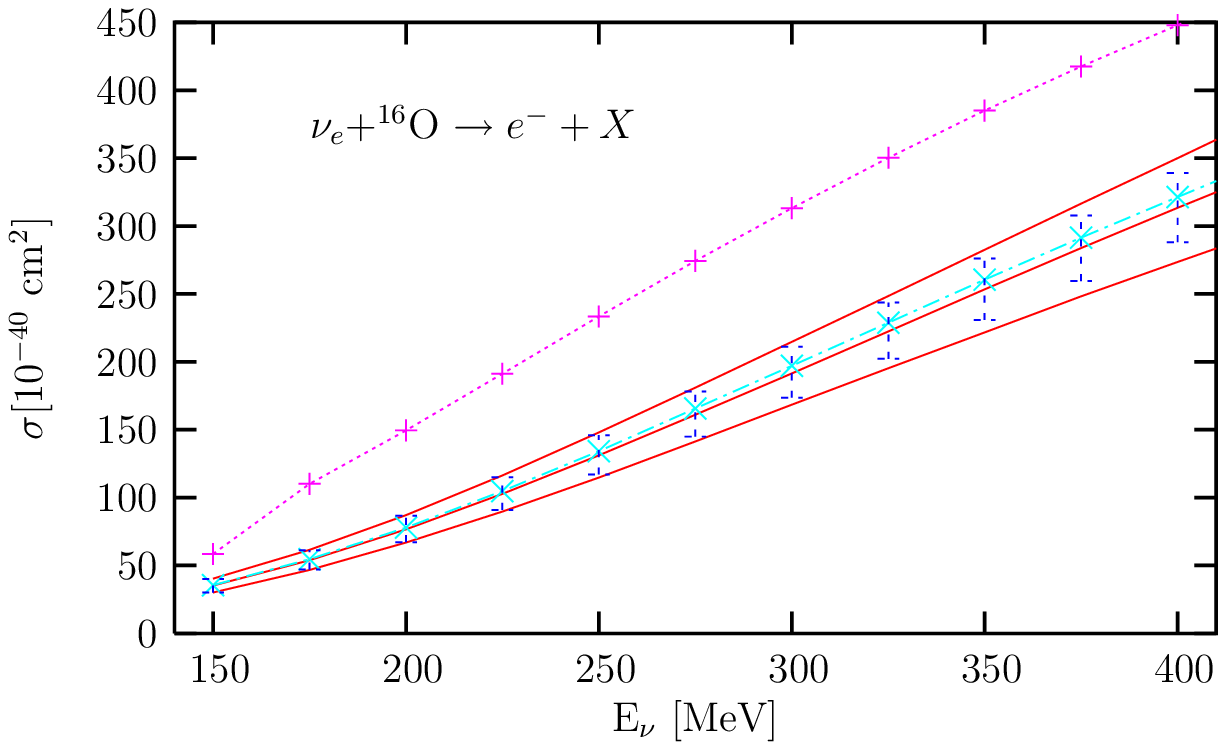}\hspace{-5cm}\includegraphics[height=20cm]{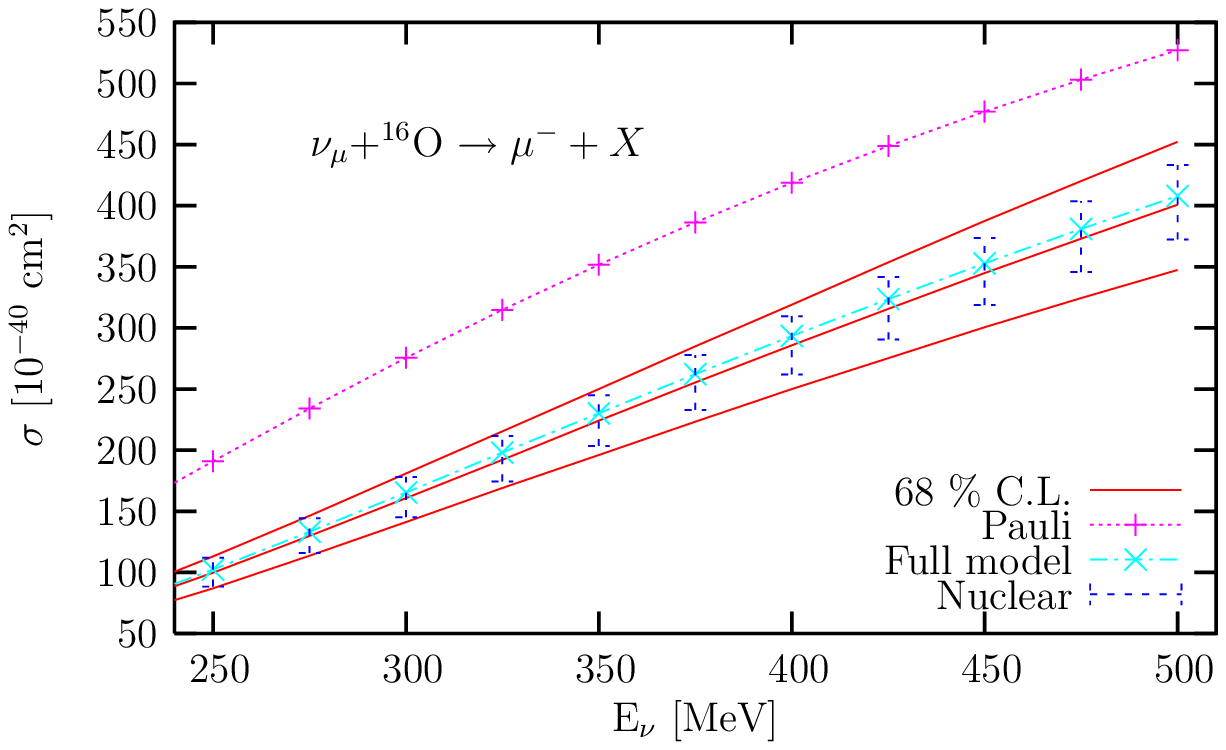}}\\[-14cm]
\makebox[0pt]{\includegraphics[height=20cm]{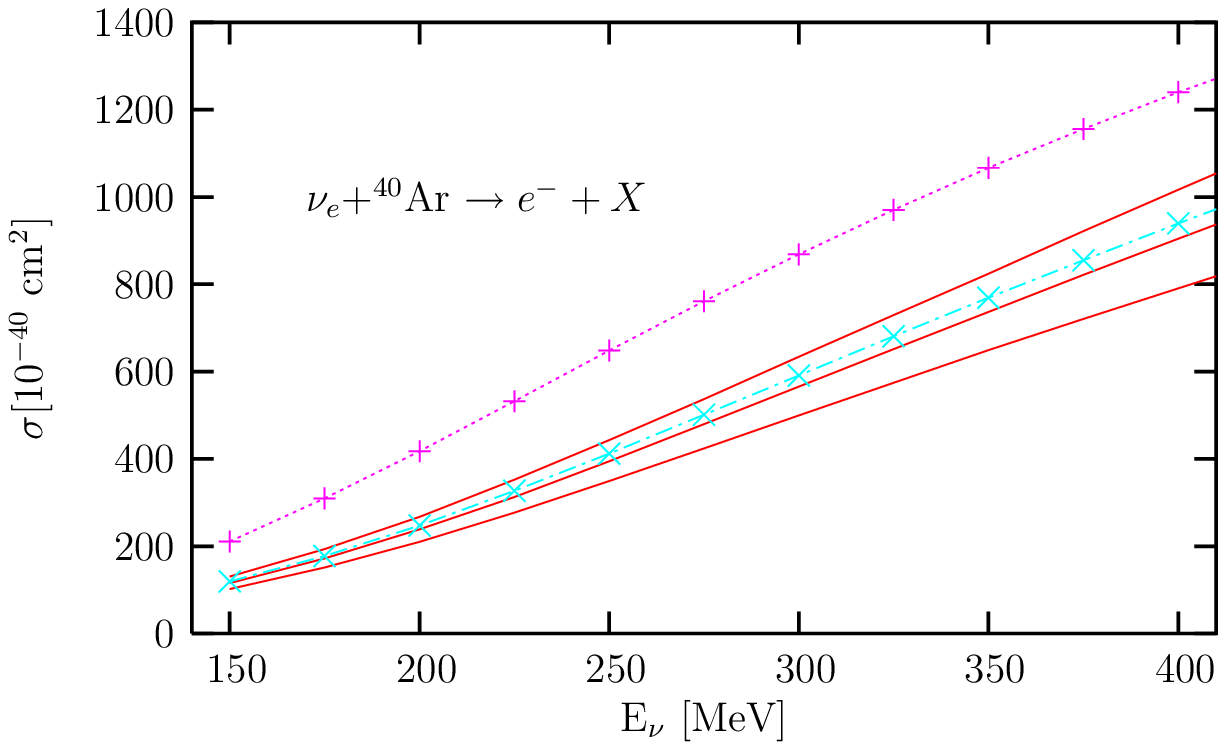}\hspace{-5cm}\includegraphics[height=20cm]{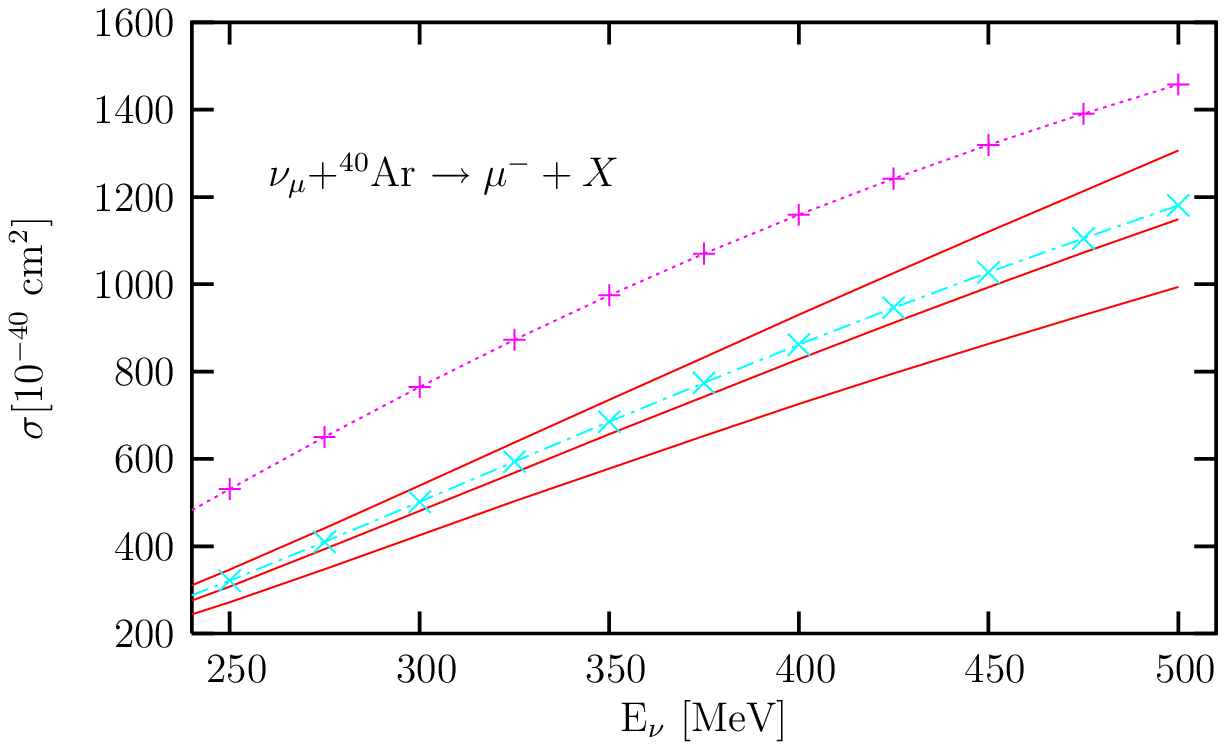}}
\end{center}
\vspace{-13.5cm}
\caption{\footnotesize (Color online) Electron and muon neutrino inclusive QE
  integrated cross sections from carbon, oxygen and argon, as a
  function of the neutrino energy. In all cases non-relativistic
  nucleon kinematics has been employed. Results denoted as ``Full
  model'' are obtained from the full model developed in
  Ref.~\protect\cite{Nie04}, while those denoted as ``Pauli'' have
  been obtained without including RPA, Coulomb and nucleon selfenergy
  effects. We also give the 68\% CL band (red or solid lines). For
  oxygen, the error bars (denoted as ``Nuclear'') account for the
  uncertainties due to the imprecise knowledge of the nucleon
  densities and of the parameters entering in the model used
  (Ref.~\protect\cite{Nie04}) to compute nuclear effects (RPA and  nucleon
  self-energy).}
\label{fig:1}
\end{figure}

\begin{figure}
\begin{center}
\makebox[0pt]{\includegraphics[height=20cm]{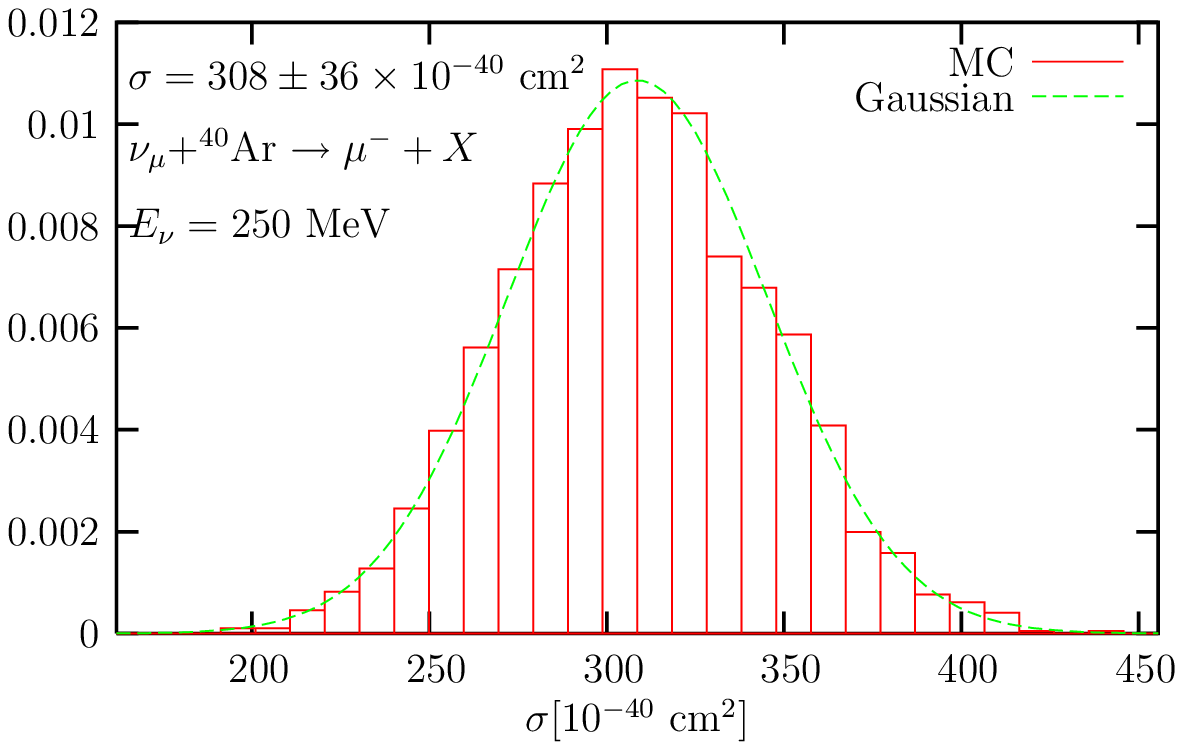}\hspace{-5cm}\includegraphics[height=20cm]{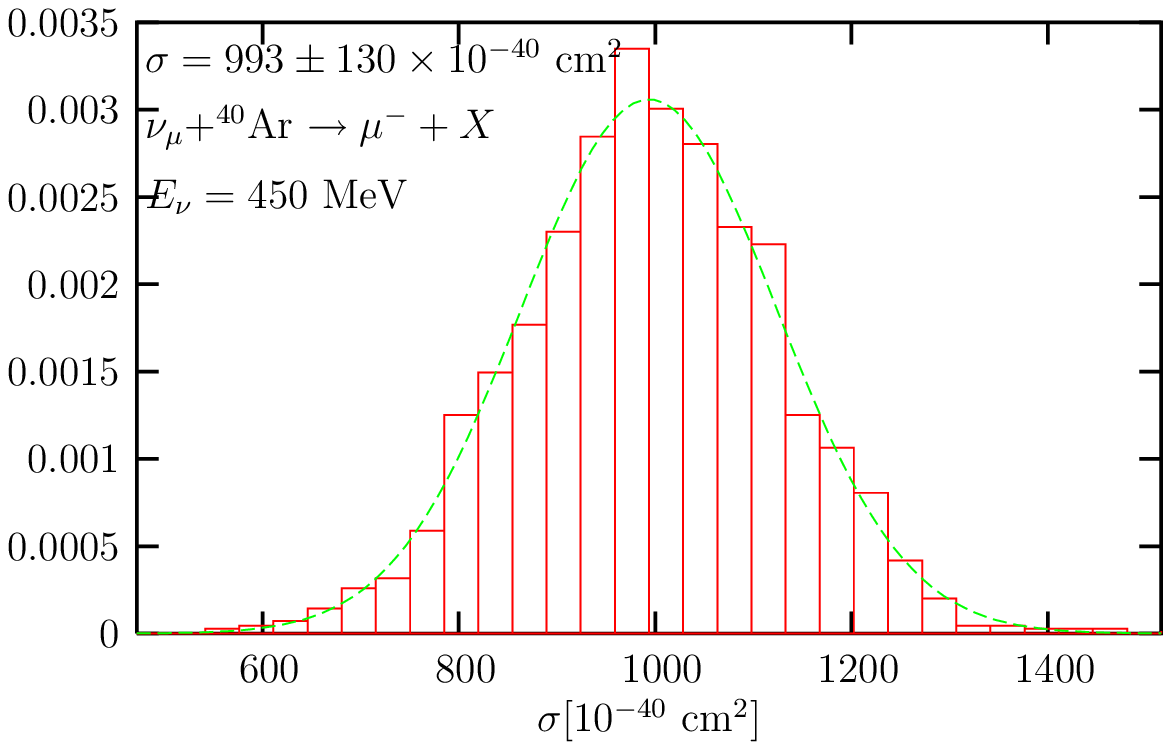}}
\end{center}
\vspace{-13.5cm}
\caption{\footnotesize (Color online) MC uncertainty distributions for muon
  neutrino inclusive QE integrated cross section from argon, at 250
  (left) and 450 (right) MeV incoming neutrino energies. The dashed 
  lines stand for Gaussian distributions with central values and variances
  indicated in each panel.}
\label{fig:2}
\end{figure}

\begin{figure}
\begin{center}
\makebox[0pt]{\includegraphics[height=30cm]{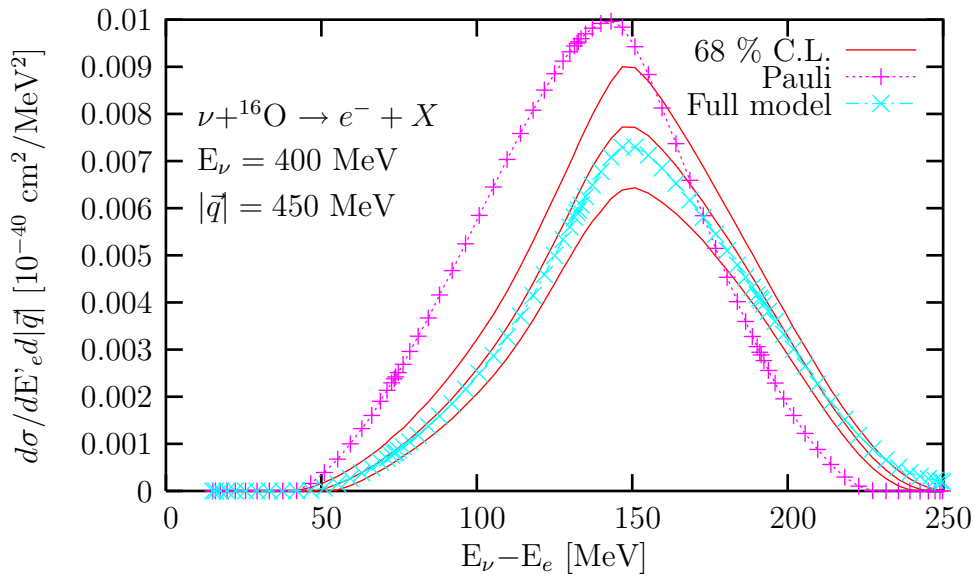}}
\end{center}
\vspace{-20.5cm}
\caption{\footnotesize (Color online) Electron neutrino inclusive QE
  differential cross section, at a fixed lepton momentum transfer of
  450 MeV, in oxygen as a function of the lepton energy transfer. The
  incoming neutrino energy is 400 MeV and non-relativistic nucleon
  kinematics has been employed. Results denoted as ``Full model'' are
  obtained from the full model developed in Ref.~\protect\cite{Nie04},
  while those denoted as ``Pauli'' have been obtained without
  including RPA, Coulomb and nucleon selfenergy effects. We also give
  the 68\% CL band (red or solid lines).}
\label{fig:3}
\end{figure}

\begin{figure}
\begin{center}
\makebox[0pt]{\includegraphics[height=25cm]{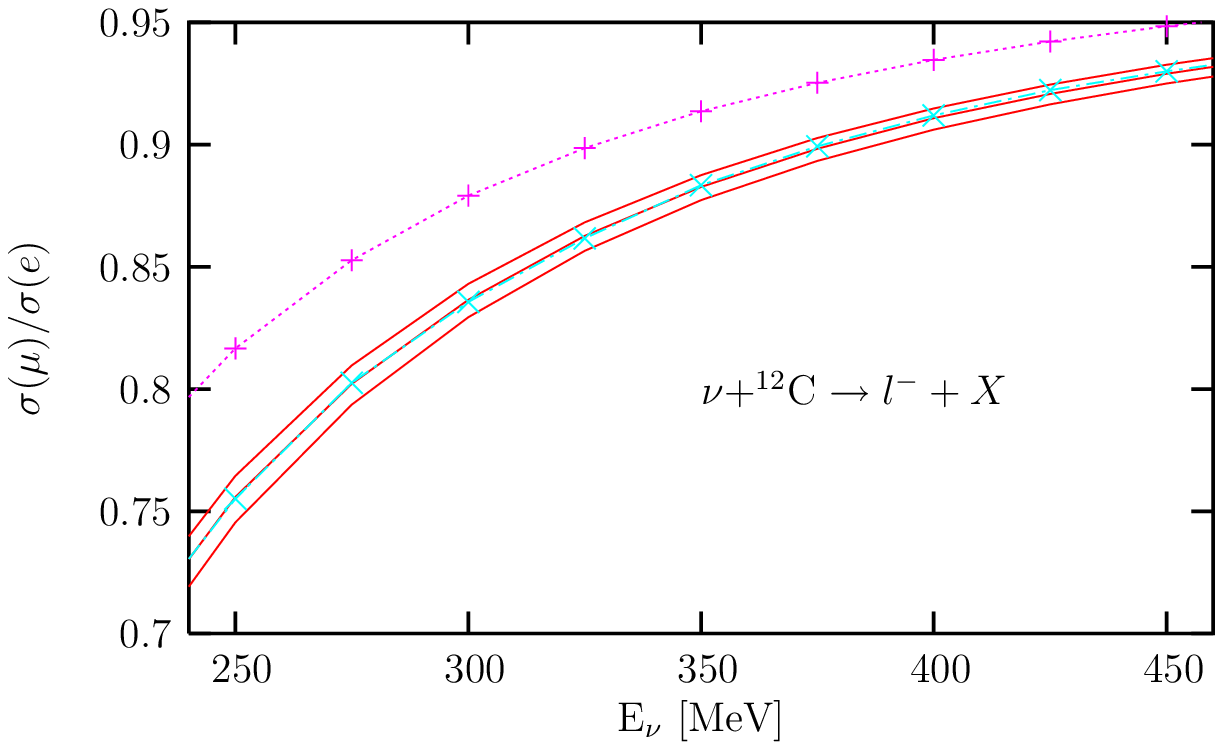}}\\[-18cm]
\makebox[0pt]{\includegraphics[height=25cm]{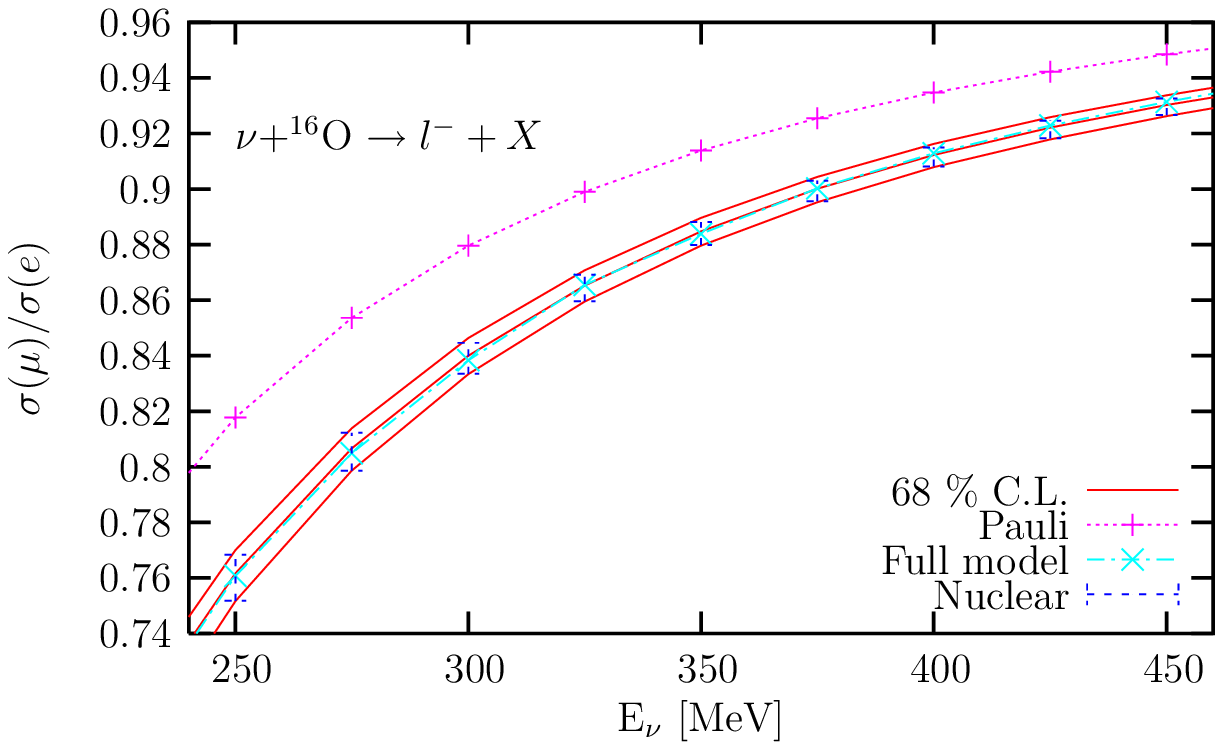}}\\[-18cm]
\makebox[0pt]{\includegraphics[height=25cm]{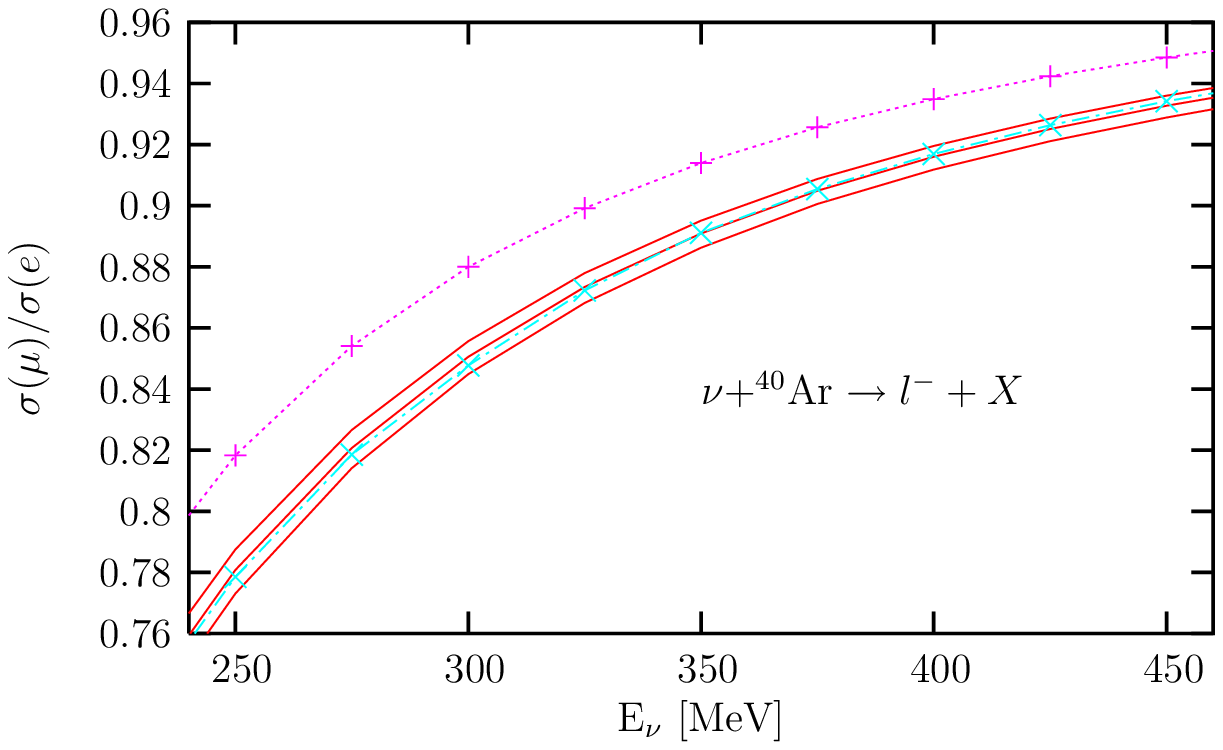}}
\end{center}
\vspace{-17cm}
\caption{\footnotesize (Color online) Ratio of inclusive QE cross sections
  $\sigma(\mu)/\sigma(e)$ for carbon, oxygen and argon, as a
  function of the incoming neutrino energy. In all cases
  non-relativistic nucleon kinematics has been employed. Besides, the
  68\% CL band (red or solid lines), we also give results (crosses) from
  the full model developed in Ref.~\protect\cite{Nie04}, and from the
  latter model without including RPA, Coulomb and nucleon selfenergy
  effects (line denoted as ``Pauli''). For oxygen, the error bars have
  the same meaning as in Figs.~\protect\ref{fig:1} and
  ~\protect\ref{fig:3}.}
\label{fig:4}
\end{figure}

\section{Results and Concluding Remarks}

 By means of a Monte Carlo simulation, we generate a total of $2000$
 sets of input parameters\footnote{We have checked that the errors
 quoted in the following are already stable when $1500$ event
 simulations are performed.}  from an uncorrelated multidimensional
 Gaussian distribution, with central values and standard deviations
 compiled in Table~\ref{tab:par}. For each of the sample sets, we
 compute the different observables discussed in this work, and thus we
 obtain the distributions of all of them. Theoretical errors and
 uncertainty bands on the derived quantities are always obtained by
 discarding the highest and lowest $16\%$ of the sample values, to
 leave a $68\%$ Confidence Level (CL) interval.

 In Figs.~\ref{fig:1} and~\ref{fig:2}, we present electron and muon
 neutrino inclusive QE integrated cross sections from carbon, oxygen
 and argon. Several comments are in order,
\begin{itemize}

\item As mentioned above, the imaginary part of the nucleon selfenergy
has been neglected when performing the MC simulation. Let us look at
Fig.~\ref{fig:1}, as can be appreciated there, the differences between
the central line\footnote{It has been obtained
with the central values of the parameters quoted in
Table~\ref{tab:par} and neglecting the imaginary part of the nucleon
selfenergy.} of the 68\% CL band and the full model prediction,
which includes the effects of the imaginary part of $\Sigma$, are
almost negligible and significantly smaller than the size of the 68\%
CL errors. This corroborates the findings of Ref.~\cite{Nie06}, and
though we will make graphically such a comparison in most of the
plots that will be presented in what follows, we will not make any
further comment about it.

\item Nuclear effects beyond implementing Pauli's exclusion principle
  and enforcing the  correct energy balance are sizeable and much
  larger than the uncertainties on the predictions deduced from the MBF
  presented in Ref.~\cite{Nie04}. 

\item In oxygen, we estimate separately the uncertainties (error
  bars\footnote{ Hence, these error bars do not take into account the
  uncertainties of the neutrino-nucleon form factors.}  in the middle
  plots of Fig.~\ref{fig:1}) due to the imprecise knowledge of the
  nucleon densities and of the parameters entering in the model used
  in Ref.~\protect\cite{Nie04} to compute nuclear effects (RPA and
  nucleon self-energy). The value of the relative  uncertainty due to
  nuclear effects    decrease with  energy, while relative errors
  induced by the neutrino-nucleon form factors increase with 
  energy, and at the higher end, they could be comparable to those
  affecting the evaluation of the nuclear effects.

\item Uncertainties on the integrated cross sections are of the order
  of 10--15\%, which turn out to be similar to those assumed for the input
  parameters (Table~\ref{tab:par}). Hence, predictions of
  Ref.~\cite{Nie04} seems stable and no fine tuning  parameters are
  identified in the model. 

\item As can be appreciated in Fig.~\ref{fig:2} for two particular
  cases, theoretical uncertainties on the
  cross section can be, in a good approximation, modeled by a
  Gaussian distribution.

\end{itemize}

Conclusions are similar for differential cross sections. As an
example, in Fig.~\ref{fig:3} we show the electron neutrino inclusive
QE double differential cross section, at a fixed lepton momentum
transfer, in oxygen as a function of the lepton energy transfer.

\begin{figure}
\begin{center}
\makebox[0pt]{\includegraphics[height=20cm]{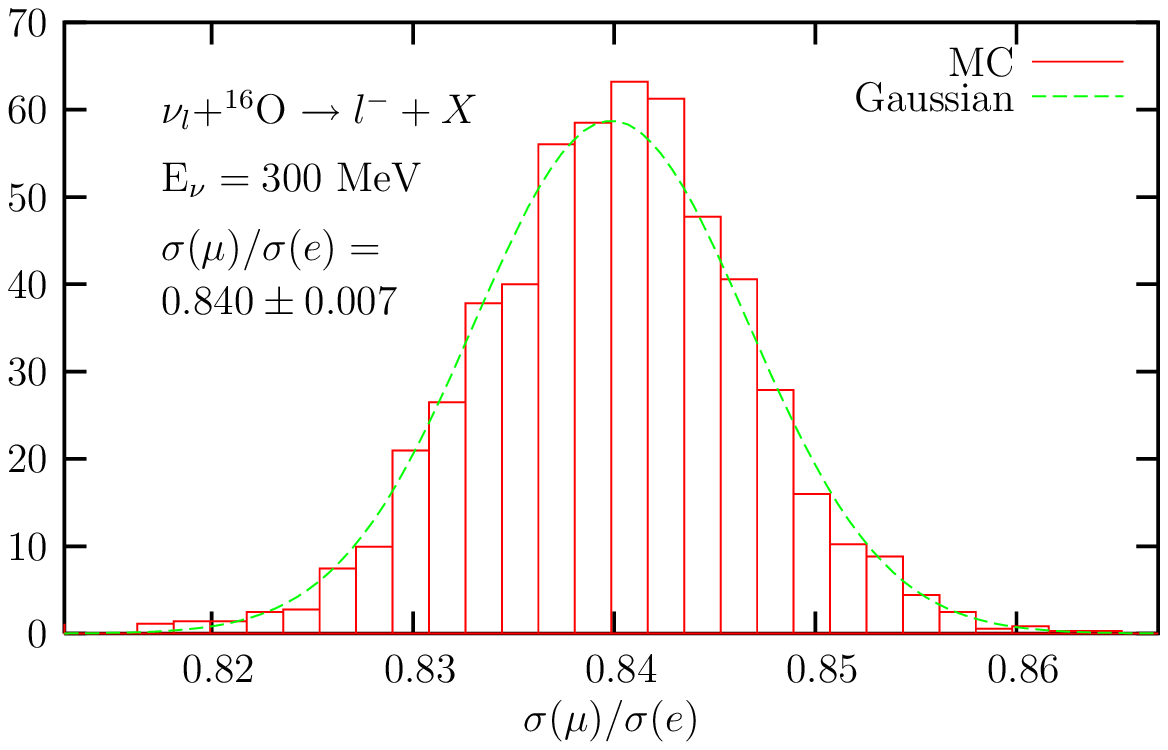}\hspace{-5cm}\includegraphics[height=20cm]{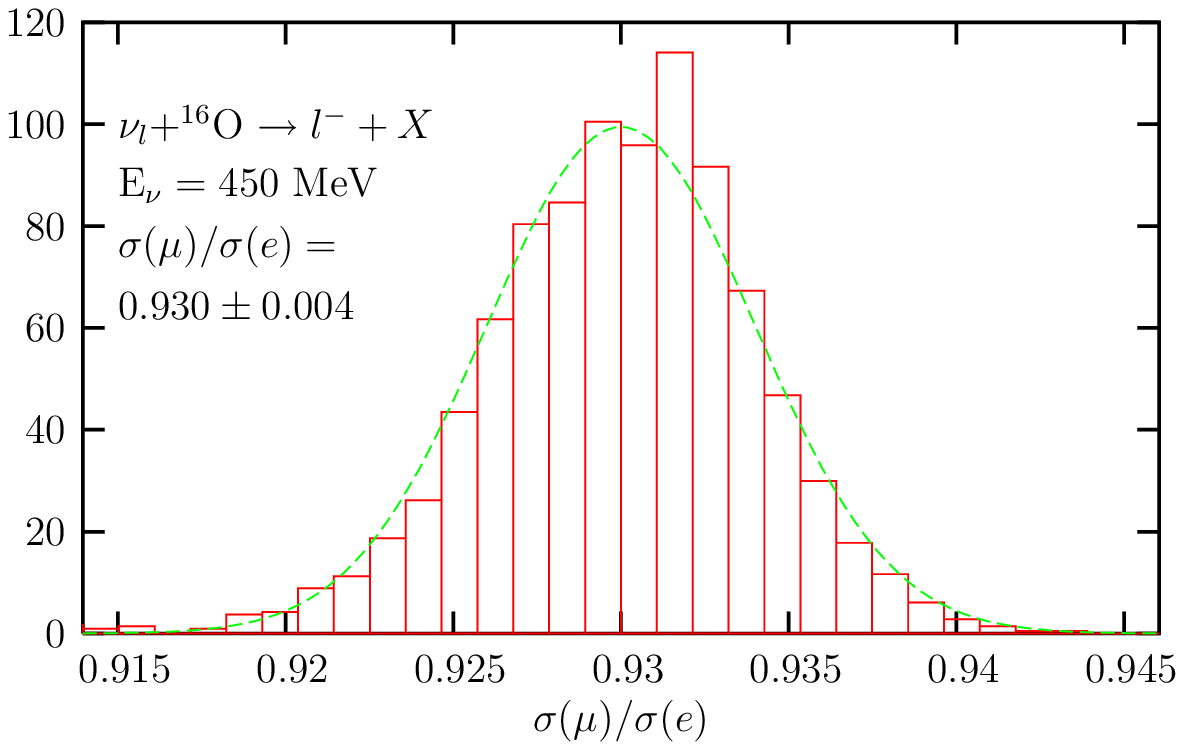}}
\end{center}
\vspace{-13.5cm}
\caption{\footnotesize (Color online) MC uncertainty distributions for
  the ratio of inclusive QE cross sections $\sigma(\mu)/\sigma(e)$
  from oxygen, at 300 (left) and 450 (right) MeV incoming neutrino
  energies. The dashed lines stand for Gaussian distributions with
  central values and variances indicated in each panel.}
\label{fig:5}
\end{figure}

Theoretical errors cancel partially out in the ratio
$\sigma(\mu)/\sigma(e) \equiv \sigma(\nu_\mu+ ^AZ \to \mu^- + X) /
\sigma(\nu_e+ ^AZ \to e^- + X) $, as can be appreciated in
Fig.~\ref{fig:4}. Theoretical uncertainties on this ratio turn out to
be smaller than 1\%. On the other hand, predictions for this ratio
obtained from a simple Lindhard function\footnote{It is to say from a
non-interacting local Fermi gas model of non-relativistic nucleons.}
incorporating a correct energy balance in the reaction (lines denoted
as ``Pauli'' in the plots) differ from the results obtained from the
model of Ref.~\cite{Nie04} only at the level of 5\%, in sharp contrast
with the situation found for each of the the individual cross sections
$\sigma(\nu_\mu+ ^AZ \to \mu^- + X)$ and $\sigma(\nu_e+ ^AZ \to e^- +
X)$ (see Fig.~\ref{fig:1}). This is consistent with the findings of a
somewhat similar analysis carried out in Refs.~\cite{En93,Mar99,
Ma93}.  Finally, here we also find that the theoretical uncertainties
on the ratio $\sigma(\mu)/\sigma(e)$ can be, in a good approximation,
modeled by a Gaussian distribution, as can be seen in
Fig.~\ref{fig:5}.

For antineutrino induced reactions we find a totally parallel
scenario to that discussed so far for neutrino--induced ones.

\begin{figure}
\begin{center}
\makebox[0pt]{\includegraphics[height=25cm]{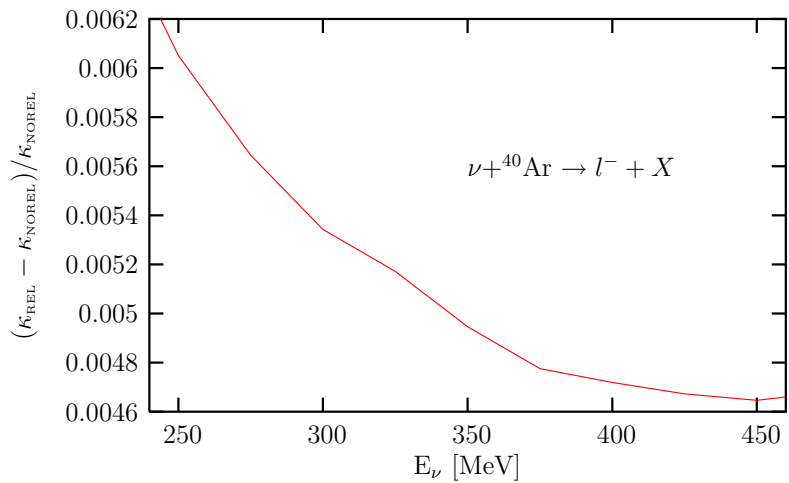}}
\end{center}
\vspace{-16.5cm}
\caption{\footnotesize (Color online) Relative corrections to the ratio
$\sigma(\mu)/\sigma(e)$, in argon, due to the use of relativistic nucleon
kinematics. We denote by $\kappa_{\rm REL}$ ($\kappa_{\rm NOREL}$)
the ratio $\sigma(\mu)/\sigma(e)$ deduced from a non-interacting local
Fermi gas model of relativistic (non-relativistic) nucleons improved
by the use of the experimental $Q-$ values (see
Ref.~\protect\cite{Nie04} for some more details). Thus, $\kappa_{\rm
  NOREL}$ would be given by the line denoted by ``Pauli'' in the argon panels
of Fig.~\protect\ref{fig:4}. }
\label{fig:6}
\end{figure}

To finish this work, we would like to discuss some other systematic
errors associated to the validity of the hypothesis in which the
scheme of Ref.~\cite{Nie04} is based, and not considered yet. In
first place, up to this point a non-relativistic nucleon kinematics has
been employed. This is because the RPA and nucleon selfenergy models
used in Ref.~\cite{Nie04} are based in a non-relativistic 
effective baryon--baryon interaction in the nuclear medium. The use of
relativistic kinematics for the nucleons leads to moderate reductions
of both neutrino and antineutrino cross sections, ranging these
reductions in the interval 4-9\%, at the intermediate energies
considered in this work. Such corrections do not depend significantly
on the considered nucleus~\cite{Nie04}. In the ratio
$\sigma(\mu)/\sigma(e)$,  relativistic nucleon kinematics effects are
quite small, being always smaller than 1\% in the whole neutrino
energy interval studied in this work, as can be seen in Fig.~\ref{fig:6}.

Second, one might think that a LFG description of the nucleus is poor,
and that a proper finite nuclei treatment is necessary. For inclusive
processes and nuclear excitation energies of around 100 MeV or higher,
the findings of Refs.~\cite{GNO97}, \cite{CO92} and~\cite{pion}
clearly contradict this conclusion. The reason is that in these
circumstances one should sum up over several nuclear configurations,
both in the discrete and in the continuum, and this inclusive sum is
almost not sensitive to the details of the nuclear wave
function\footnote{The results of Ref.~\cite{Nie04} for the inclusive
muon capture in nuclei through the whole periodic table, where the
capture widths vary from about 4$\times 10^4$ s$^{-1}$ in $^{12}$C to
1300 $\times 10^4$ s$^{-1}$ in $^{208}$Pb, and of the LSND
measurements of the $^{12}$C $(\nu_\mu,\mu^-)X$ and $^{12}$C
$(\nu_e,e^-)X$ reactions near threshold indicate that the predictions
of our scheme, for totally integrated inclusive observables, could
even be extended to much smaller, of the order of 10 or 20
MeV, nuclear excitation energies.  In this respect, the works of
Refs.~\cite{mucap} and ~\cite{picap} for inclusive muon capture and
radiative pion capture in nuclei, respectively, turn out to be quite
enlightening. In those works, continuum shell model results are
compared to those obtained from a LFG model for several nuclei from
$^{12}$C to $^{208}$Pb. The differential decay width shapes predicted
for the two set of models are substantially different. Shell model
distributions present discrete contributions and in the continuum
appear sharp scattering resonances. Despite the fact that those
distinctive features do not appear in the LFG differential decay
widths, the totally integrated widths (inclusive observable) obtained
from both descriptions of the process do not differ in more than 5 or
10\%. The typical nuclear excitation energies in muon and radiative
pion capture in nuclei are small, of the order of 20 MeV, and thus one
expects that at higher excitation energies, where one should sum up
over a larger number of nuclear final states, the LFG predictions for
inclusive observables would become even more reliable.}, in sharp
contrast to what happens in the case of exclusive processes where the
final nucleus is left in a determined nuclear level. On the other
hand, the LFG description of the nucleus allows for an accurate
treatment of the dynamics of the elementary processes (interaction of
gauge bosons with nucleons, nucleon resonances, and mesons,
interaction between nucleons or between mesons and nucleons, etc.)
which occur inside the nuclear medium. Within a finite nuclei
scenario, such a treatment becomes hard to implement, and often the
dynamics is simplified in order to deal with more elaborated nuclear
wave functions. This simplification of the dynamics cannot lead to a
good description of nuclear inclusive electroweak processes at the
intermediate energies of interest for future neutrino experiments.

For all of this, it is sound to assume relative errors of about
10-15\% on the QE neutrino--nucleus (differential and integrated)
cross sections predicted by the model of Ref.~\cite{Nie04}, at
intermediate energies. Uncertainties on the ratio
$\sigma(\mu)/\sigma(e)$ would be certainly smaller, likely not larger
than about 5\%, and mostly coming from deficiencies of the LFG picture
of the nucleus assumed in Ref.~\cite{Nie04}.

\begin{table}
\begin{tabular}{rcrcl rcrcl}\hline\hline
\multicolumn{5}{c}{Form Factors} & \multicolumn{5}{c}{Nucleon Interaction}\\
\hline
$M_D$ & = & 0.843 & $\pm$ &0.042 GeV &$f_{0}^{\prime (in)}$ & = & 0.33
& $\pm$ & 0.03\\ 
$\lambda_n$ & = & 5.6 & $\pm$ & 0.6 &$f_{0}^{(\prime ex)}$ & = & 0.45
 & $\pm$ &0.05\\ 
$M_A$ & = & 1.05 & $\pm$ & 0.14 GeV &$f$ & = & 1.00 & $\pm$ &
0.10  \\
  $g_A$ & = & 1.26 & $\pm$ & 0.01 & $f^{*}$   & = & 2.13 & $\pm$ & 0.21  \\   
 & & & & & $\Lambda_\pi$      & = & 1200 & $\pm$ & 120 MeV  \\
 & & & & & $C_\rho$           & = & 2.0    & $\pm$ & 0.2      \\
 & & & & & $\Lambda_\rho$     & = & 2500 & $\pm$ & 250 MeV  \\
 & & & & & $g^\prime$         & = & 0.63 & $\pm$ & 0.06     \\ \hline\hline
\end{tabular}
\caption{Central values and errors of the model input parameters,
    besides we have also included 10\% uncertainties (relative) in
    both the real part of the nucleon selfenergy and densities (see
    text for details).}\label{tab:par}
\end{table}

\begin{acknowledgments}

J.N. warmly thanks to M. J. Vicente-Vacas for various stimulating
discussions and communications. This work was supported by DGI and
FEDER funds, contract BFM2005-00810, by the EU Integrated
Infrastructure Initiative Hadron Physics Project contract
RII3-CT-2004-506078 and by the Junta de Andaluc\'\i a (FQM-225).

\end{acknowledgments}


\begin{thebibliography}{blabla}


\bibitem{fukuda} Super-Kamiokande Collaboration (Y. Fukuda et al.,)
  Phys. Rev. Lett. {\bf 81} (1998) 1562; Y.~Fukuda {\it et al.},
  Phys.\ Lett.\ B {\bf 433}, 9 (1998).


\bibitem{procs} Proceedings of the RCCN International Workshop On
        Sub-dominant Oscillation Effects in Atmospheric Neutrino
        Experiments Kashiwa, 2004, Eds. T. Kajita and K. Okumura,
        Universal Academy Press, Inc.\ Tokyo, Japan

\bibitem{kajita}T.~Kajita and Y.~Totsuka, 
Rev.\ Mod.\ Phys.\ {\bf 73}, 85 (2001).




\bibitem{HT00} A.C. Hayes and I.S. Towner, Phys. Rev. {\bf C61} (2000) 044603.


\bibitem{Au02} N. Auerbach, B.A. Brown,  Phys. Rev.{\bf C65} (2002) 024322.

\bibitem{Ja02} N. Jachowicz, K. Heyde, J. Ryckebusch and S. Rombouts,
  Phys. Rev. {\bf C65} (2002) 025501. 

\bibitem{Ko03} E. Kolbe, K. Langanke, G. Mart\'\i
 nez-Pinedo and P.Vogel, J. Phys. {\bf G29} (2003) 2569.


\bibitem{Ko97} E. Kolbe, K. Langanke
 and P. Vogel, Nucl. Phys. {\bf A652} (1999) 91.

\bibitem{Au97}  N. Auerbach, N. Van Giai and O.K. Vorov,
  Phys. Rev. {\bf C56} (1997) R2368; C. Volpe, et al., Phys. Rev. {\bf
  C62} (2000) 015501; N. Auerbach et al., Nucl. Phys. {\bf A687}
  (2001) 289c.

\bibitem{Bl01} C. Bleve, et al., Astr. Part. Phys. {\bf 16} (2001)
  145.

\bibitem{Ma03} C. Maieron, M.C. Martinez, J.A. Caballero and
J.M. Udias, Phys. Rev. {\bf C68} (2003) 048501. 

\bibitem{MGP03} A. Meucci, C. Giusti and F.D. Pacati, Nucl. Phys.~{\bf
  A739} (2004) 277.

\bibitem{Amaro:2004bs}
  J.E. Amaro, M.B. Barbaro, J.A. Caballero, T.W. Donnelly, A. Molinari 
and I. Sick, Phys. Rev.  {\bf C71} (2005) 015501;  J.A. Caballero,
  J.E. Amaro, M.B. Barbaro, T.W. Donnelly, C. Maieron and J.M. Udias,
  Phys. Rev. Lett.  {\bf 95} (2005) 252502. 


\bibitem{Nie04} J.~Nieves, J.~E.~Amaro, and M.~Valverde,
                Phys.\ Rev.\ C {\bf 70}, 055503 (2004) [Erratum-ibid.\
                C {\bf 72}, 019902 (2005)].

\bibitem{Nie06} J.~Nieves,  M.~Valverde and M.J. Vicente-Vacas,
                Phys.\ Rev.\ C {\bf 73} 025504 (2006). 


\bibitem{GNO97} A. Gil, J. Nieves and E. Oset, Nucl. Phys. {\bf A627} (1997)
  543; {\it ibidem}, Nucl. Phys. {\bf A627} (1997) 599. 

\bibitem{CO92} R.C. Carrasco and E. Oset, Nucl. Phys. {\bf A536}
  (1992) 445.

\bibitem{OTW82} E. Oset, H. Toki and W. Weise, Phys. Rep.  {\bf 83} (1982) 281.

\bibitem{pion} L.L. Salcedo, E. Oset, M.J. Vicente-Vacas and
C. Garc\'\i a Recio, Nucl. Phys. {\bf A484} (1988) 557; J. Nieves,
E. Oset, C. Garc\'\i a-Recio, Nucl. Phys. {\bf A554} (1993) 509; {\it
ibidem} Nucl. Phys. {\bf A554} (1993) 554; J. Nieves and E. Oset,
Phys. Rev. {\bf C47} (1993) 1478; E. Oset, P. Fern\'andez de
C\'ordoba, J. Nieves, A. Ramos and L.L. Salcedo,
Prog. Theor. Phys. Suppl. {\bf 117} (1994) 461; C. Albertus,
J.E. Amaro and J. Nieves, Phys. Rev. Lett. {\bf 89} (2002) 032501;
{\it ibidem} Phys. Rev. {\bf C67} (2003) 034604.



\bibitem{Galster}   S.~Galster, H.~Klein, J.~Moritz, K.~H.~Schmidt,
                 D.~Wegener, and J.~Bleckwenn,
                 Nucl.\ Phys.\ {\bf B32} 221 (1971). 

\bibitem{ADW} V. Bernard, L. Elouardrhiri, U.-G. Meissner,
  J. Phys. {\bf G28} (2002) R1; D. Barquilla-Cano, A.J. Buchmann,
  E. Hern\'andez, Nucl. Phys. {\bf A714} (2003) 611.

\bibitem{pdg} S. Eidelman, {\it et al.}, Phys. Lett. {\bf B592} (2004) 1.

\bibitem{BNL1} N. J. Baker {\it et al.,} Phys. Rev {\bf D23} (1981)
  2499. 

\bibitem{BNL2} T. Kitagaki {\it et al.,} Phys. Rev {\bf D42} (1990) 1331. 

\bibitem{speth}  J.~Speth, E.~Werner, and W.~Wild,
                Phys.\ Rep.\ {\bf 33} 127 (1977); 
                J.~Speth, V.~Klemt, J.~Wambach, and G.~E.~Brown 
                Nucl.\ Phys.\ {\bf A343} 382 (1980).

\bibitem{FO92} P. Fern\'{a}ndez de C\'{o}rdoba and E. Oset,
  Phys. Rev. {\bf C46} (1992) 1697.

\bibitem{Fa83} S. Fantoni, B.L. Friman, and V.R. Pandharipande,
  Nuc. Phys. {\bf A399} (1983) 51; S. Fantoni and V.R. Pandharipande,
  Nucl. Phys. {\bf A427} (1984) 473.

\bibitem{Ma85}  C. Mahaux, P.F. Bortignon, R.A. Broglia, and
  C.H. Dasso, Phys. Rep. {\bf 120} (1985) 1.

\bibitem{Ra89} A. Ramos, A. Polls, and W, H. Dickhoff,
  Nucl. Phys. {\bf A503} (1989) 1

\bibitem{Mu95} H. M\"uther, G. Knehr and A. Polls, Phys. Rev. {\bf
  C52} (1995) 2955.

\bibitem{ExpDens} C.~W.~de Jager, H.~de Vries, and C.~de Vries,
    At.\ Data and Nucl.\ Data Tables {\bf 14} 479 (1974); {\bf 36} 495 (1987).

\bibitem{Ne75} J.W. Negele and D. Vautherin, Phys. Rev. {\bf C11}
(1975) 1031 and references therein.

\bibitem{GNO92} C. Garc\'\i a-Recio, J. Nieves and 
E. Oset, Nucl. Phys. {\bf A547} (1992) 473.

\bibitem{En93} J. Engel, E. Kolbe, K. Langanke and P. Vogel,
Phys. Rev. {\bf D48} (1993) 3048.

\bibitem{Mar99} J. Marteau, Eur. Phys. Jour. {\bf A5} (1999) 183.

\bibitem{Ma93} A. K. Mann, Phys. Rev. {\bf D48} (1993) 422.


\bibitem{mucap} J.E. Amaro, C. Maieron, J. Nieves and
M. Valverde,  Eur. Phys. Jour. {\bf A24} (2005) 343 
[Erratum-ibid.\  {\bf A26}, 307 (2005)].

\bibitem{picap} J.E. Amaro, A.M. Lallena and J. Nieves, 
 Nucl. Phys. {\bf A623} (1997) 529. 



\end{thebibliography}
\end{document}